\documentclass[journal]{IEEEtran}

\usepackage{xspace,amsmath,amssymb,amsfonts, epsfig,syntonly,mathtools}
\usepackage{cite,bbm,bm,color,url, textcomp, empheq,boxedminipage}
\usepackage{algorithmicx,algorithm,algpseudocode}
\usepackage{epstopdf}
\usepackage{empheq}
\usepackage{graphicx,graphics,subfigure}
\usepackage{multirow,multicol}
\usepackage{psfrag}
\usepackage{stfloats}
\usepackage{url}
\newtheorem{definition}{Definition}

\long\def\symbolfootnote[#1]#2{\begingroup
\def\thefootnote{\fnsymbol{footnote}}
\footnote[#1]{#2}\endgroup}

\psfull

\usepackage{ifpdf}
 \ifpdf
 \else
 \fi
\begin{document}

\title{Joint Task Offloading and Cache Placement for Energy-Efficient Mobile Edge Computing Systems}

\author{Jingxuan Liang, Hong Xing,~\IEEEmembership{Member, IEEE}, Feng Wang,~\IEEEmembership{Member, IEEE}, and Vincent K. N. Lau, \IEEEmembership{Fellow, IEEE}\\
 \thanks{J. Liang and F. Wang are with the School of Information Engineering, Guangdong University of Technology, Guangzhou 510006, China (e-mail: fengwang13@gdut.edu.cn).{\em (Corresponding author: Feng Wang)}}
 \thanks{H. Xing is with Internet of Things Thrust, The Hong Kong University of Science and Technology (Guangzhou), Guangzhou 511453, China, and is also with the Department of ECE, The Hong Kong University of Science and Technology, Hong Kong (e-mail: hongxing@ust.hk).}
 \thanks{V. K. N. Lau is with the Department of ECE, The Hong Kong University of Science and Technology, Hong Kong (e-mail: eeknlau@ust.hk).}

 \vspace{-1.2cm}
 }

\maketitle

\begin{abstract}
 This letter investigates a cache-enabled multiuser mobile edge computing (MEC) system with dynamic task arrivals, taking into account the impact of proactive cache placement on the system's overall energy consumption. We consider that an access point (AP) schedules a wireless device (WD) to offload computational tasks while executing the tasks of a finite library in the {\em task caching} phase, such that the nearby WDs with the same task request arriving later can directly download the task results in the {\em task arrival and execution} phase. We aim for minimizing the system's weighted-sum energy over a finite-time horizon, by jointly optimizing the task caching decision and the MEC execution of the AP, and local computing as well as task offloading of the WDs at each time slot, subject to caching capacity, task causality, and completion deadline constraints. The formulated design problem is a mixed-integer nonlinear program. Under the assumption of fully predicable task arrivals, we first propose a branch-and-bound (BnB) based method to obtain the optimal offline solution. Next, we propose two low-complexity schemes based on convex relaxation and task-popularity, respectively. Finally, numerical results show the benefit of the proposed schemes over existing benchmark schemes.
 \end{abstract}	


\vspace{-0.5cm}
\section{Introduction}
 Various computation-extensive internet of things (IoT) applications (such as extended reality, auto-driving, and tactile networks) call for low-latency communication and computation~\cite{ref1}. By deploying dedicated edge servers at the network edge, mobile edge computing (MEC) has been recognized as an enabling technology to meet the stringent requirement of these delay-sensitive services while addressing the computation/communication resource limitation issue of these wireless devices (WDs)\cite{FW18, FW20, AL19}. Leveraging the storage resources of the MEC servers to proactively cache computational tasks for possible reuse, the computation performance of the MEC system can be further enhanced.

 Compared to the conventional MEC system designs without caching capabilities\cite{FW18, FW20, AL19}, cache-enabled MEC system designs encounter several new technical challenges. First, the task caching and offloading decisions need to be jointly made so as to make the best use of the limited caching and computation resources. Second, the task caching and offloading strategies need to be adaptive to task dynamics and the WDs' mobility. Finally, to improve the energy efficiency of the cache-enabled MEC system, it is imperative to jointly optimize the system's computation, caching, and communication resources. In the literature, there exist several works investigating cache-enabled MEC system designs\cite{GUO18, YAN21, WANG17, XU18, S10}. For example, an adaptive task offloading and caching scheme was proposed to provide high-quality video services to vehicular users\cite{GUO18}. Based on a two-stage dynamic game strategy,  \cite{WANG17,YAN21} investigated joint computation offloading and resource allocation design for cache-enabled MEC systems. The works \cite{XU18} and \cite{S10} proposed the joint service caching and task offloading design in the dense cellular network and single-user scenarios, respectively. Note that most of the above existing works \cite{GUO18, YAN21, WANG17, XU18, S10} failed to consider the benefit of proactive caching to the overall multiuser MEC systems with WDs' dynamical task arrivals over time slots.

 In this letter, we investigate an energy-efficient cache-enabled multiuser MEC system with dynamic task arrivals over a finite-time horizon. The finite-time horizon consists of the {\em task caching} phase and the {\em task arrival and execution} phase. We consider that the MEC server selects and proactively caches the result of  several tasks from a finite library in the task caching phase; at the task arrival and execution phase, the WDs can directly download the task results if their requested tasks have been cached by the MEC server, and perform local computing and task offloading otherwise. We jointly optimize the task cache placement decision and remote computing of the AP, and task offloading as well as local computing of the WDs at each time slot, so as to minimize the system's weighted-sum energy consumption over the horizon. For obtaining a lower-bound benchmark for practical design schemes with dynamic task arrivals but imperfect prediction, we assume that the computational task sequence of each WD is fully predictable. We employ the branch-and-bound (BnB) method to obtain the optimal offline solution. Next, to facilitate the cache-enabled MEC design with low computational complexity, we propose a convex-relaxation based scheme and a task-popularity based scheme, respectively. Finally, numerical results show the benefits of our proposed schemes over existing benchmarks.

\vspace{-0.2cm}
 \section{System Model and Problem Formulation}
\begin{figure}
 \centering
\includegraphics[width=2.5in]{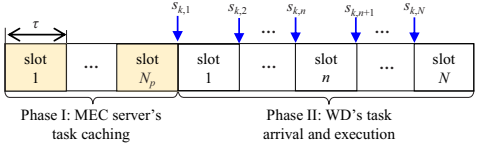}
 \caption{Timeline of the cache-enabled MEC protocol within one block.}\label{fig:protocol}
 \end{figure}

 We consider a cache-enabled multiuser MEC system, which consists of an AP (integrated with an MEC server) and a set $\mathcal {K}\triangleq \{1,..., K\}$ of single-antenna WDs. These $K$ WDs need to compute the randomly arrived tasks within a given completion deadline. Denote by ${\cal T}\triangleq\{T_1, T_2,..., T_L\}$ the computational task set to be possibly processed by the $K$ WDs. We consider a block-by-block cache-enabled MEC system, where each transmission block is divided into Phase I which is {\em MEC server's task caching} and Phase II which is {\em WDs' task arrival and execution}. Phase I and phase II are, respectively, further composed of $N_p$ and $N$ equal-duration time slots each with length $\tau$. Without loss of generality, we focus on cache-enabled MEC design within one block as shown in Fig.~\ref{fig:protocol}. To guarantee a high efficiency of this cache-enabled MEC system, it is assumed that $N_p<N$. For the tasks which are not cached at the AP, the WDs need to perform local computing and/or to offload the tasks to the MEC server for remote execution, i.e., task offloading.

 \subsection{Phase I: MEC Server's Task Caching}
 \subsubsection{Task Cache Placement}Let $\alpha_{\ell}\in\{0,1\}$ denote the caching decision for task $T_{\ell}$ at the MEC server, where the task $T_{\ell}$ is cached if $\alpha_{\ell}=1$, and $\alpha_{\ell}=0$ otherwise, $\forall \ell=1,..., L$\footnote{The caching decision variables $\{\alpha_\ell\}_{\ell=1}^L$ will be specified by the solution to an optimization problem (P1) detailed in Section III.}. By denoting $D^{\max}$ the caching capacity of the MEC server, the MEC caching needs to satisfy
 \begin{equation}\label{storage_cons1}
 \sum_{\ell= 1}^L \alpha_{\ell} D_{\ell} \leq D^{\max},
 \end{equation}
 where $D_\ell$ denotes the number of input-bits for task $T_\ell$.

 \subsubsection{Task Offloading for Caching} For facilitating the cache-enabled multiuser MEC system design, we consider the MEC server's cached tasks are all generated and offloaded from one selected WD with the smallest pathloss for task offloading to the AP. Denote by WD-$k_o$ the selected WD, where $k_o\in{\cal K}$. In order to spare the MEC server sufficient time to execute the cached tasks in the task caching phase, WD-$k_o$ needs to fully offload a number $\sum_{\ell=1}^L \alpha_{\ell}D_{\ell}$ of task input-bits by the end of the $(N_p-1)$th slot. Within the task caching phase, denote by $\tilde{d}^{\text{off}}_{k_o, i}$ the number of task input-bits offloaded from WD-$k_o$ to the AP at the $i$th slot, where $i=1,..., N_p-1$. Hence, we have
 \begin{equation}\label{storage_cons2}
 \sum_{i=1}^{N_p-1} \tilde{d}^{\text{off}}_{k_o, i} = \sum_{\ell=1}^L \alpha_{\ell} D_{\ell}.
 \end{equation}

 During the task caching phase, the amount of energy consumption of WD-$k_o$ due to task offloading is $\tilde{E}^{\text{off}}_{k_o} = \sum_{i=1}^{N_p-1} \frac{\tau\sigma^2(2^{{\tilde{d}_{k_o, i}^{\text{off}}}/(\tau B_{k_o, i})}-1)}{|h_{k_o, i}|^2}$, where $h_{k_o, i}$ and $B_{k_o, i}$ denote the complex-valued channel coefficient and system bandwidth for task offloading from WD-$k_o$ to the AP at the $i$th slot of the task caching phase, respectively, and $\sigma^2$ denotes the additive white Gaussian noise (AWGN) power at the AP receiver.

 \subsubsection{Cached Task Execution}
  The AP executes all cached tasks to proactively obtain their results for further reuse. Due to the causality of task execution, the total number of task input-bits to be executed by the MEC server until the $i$th slot of the task caching phase {\em cannot} exceed that offloaded by WD-$k_o$ before the $(i-1)$th slot, where $i=1,..., N_p$. Denote by $\tilde{d}^{\text{mec}}_i$ the number of task input-bits executed by the MEC server at the $i$th slot of the task caching phase. Accordingly, the task causality constraints in the task caching phase are
 \begin{equation}\label{eq:task-causality-for-caching}
  \sum_{j=1}^{i} \tilde{d}^{\text{mec}}_j \leq \sum_{j=1}^{i-1} \tilde{d}^{\text{off}}_{k_o, j},~\forall i=1,..., N_p,
 \end{equation}
 where $\tilde{d}_1^{\text{mec}}=0$ due to the fact that there exists no task to execute yet at the first slot of the task caching phase. In addition, the computation of the offloaded tasks needs to be completed by the MEC server within the task caching phase. Hence, we have the task completion constraint as
 \begin{equation}\label{eq:task-ddl-for-caching}
 \sum_{j=1}^{N_p} \tilde{d}^{\text{mec}}_j = \sum_{j=1}^{N_p-1} \tilde{d}^{\text{off}}_{k_o, j}.
 \end{equation}

 In addition, the amount of energy consumption of the MEC server within the task caching phase is $\tilde{E}^{\text{mec}} = \sum_{i=1}^{N_p} \zeta_0 C_0\tilde{d}_i^{\text{mec}} (\tilde{f}_i^{\text{mec}})^2=\sum_{i=1}^{N_p}\frac{\zeta_0C_0^3(\tilde{d}_i^{\text{mec}})^3}{\tau^2}$, where $\tilde{f}_i^{\text{mec}}=\frac{C_0\tilde{d}_i^{\text{mec}}}{\tau}$ denotes the required CPU rate for task execution by the MEC server at the $i$th slot of Phase I, $C_0$ denotes the number of required CPU cycles per task input-bit, and $\zeta_0$ denotes the CPU architecture capacitance coefficient of the MEC server.

\vspace{-0.2cm}
\subsection{Phase II: WDs' Task Arrival and Execution}
Within this phase, if the results of the task arriving at the beginning of a slot for WD-$k$ has been cached by the MEC server during Phase I, WD-$k$ will download the results directly\footnote{We assume that the number of task-output bits is significantly smaller than that of task-input bits, and therefore the incurred energy cost at the MEC server is negligible\cite{FW18,ref1,FW20}.}. Otherwise, this task needs to be executed by local computing at WD-$k$ and/or task offloading to the MEC server. Let $\bm s_k\triangleq\{s_{k,1},..., s_{k,N}\}$ denote the sequence of computation tasks for each WD-$k$, where each task $s_{k,n}\in{\cal T}$ arrives at WD-$k$ at the beginning of the $n$th slot and $n\in{\cal N}\triangleq \{1,...,N\}$.\footnote{We assume that the sequence of each WD's computational tasks is fully predicted by exploiting the historical data \emph{a priori}\cite{AL19,GUO18,YAN21}. Hence, the proposed solution is offline, serving as a performance lower bound for online solutions considering (partially) unknown dynamic dynamic task arrivals.} Since the arrived tasks are randomly sampled from the task set ${\cal T}$, it is possible that some tasks in the sequence $\bm s_{k}$ may be repeated. Therefore, we need to retrieve the task-arrival set from each WD-$k$'s task sequence $\bm s_k$.

 \begin{definition}[Causality Task Set]\label{def:CTS}
 For each WD-$k$, we define ${\cal S}^{\text{CTS}}_{k,n}=\{s_{k, i}\in\mathcal{T} \mid i\in\{1,...,n\}\}$ as WD-$k$'s causality task set (CTS) till the $n$th slot. It follows that ${\cal S}^{\text{CTS}}_{k,1}=\{s_{k,1}\}$ and ${\cal S}^{\text{CTS}}_{k,i}\subseteq{\cal S}^{\text{CTS}}_{k, j}$ for $i<j\in{\cal N}$.
 \end{definition}

 We consider {\em partial offloading} policy\cite{FW20R}, such that each WD-$k$ can arbitrarily divide each task into two parts for local computing and computation offloading, respectively.
\subsubsection{Local Computing and Task Offloading of WDs}
 Let $d_{k, n}^{\text{loc}} \geq 0$ and $d_{k, n}^{\text{off}} \geq 0$ denote the number of task input-bits for local computing and computation offloading for each WD-$k$ at the $n$th slot, respectively. For WD-$k$, the total number of task input-bits executed by both local computing and offloading until the $n$th slot must be smaller than those arriving until the $n$th slot, where $n\in{\cal N}$. Therefore, we have the task computation causality constraints as \cite{FW20R}
 \begin{equation}\label{eq.Task-Causality}
 \sum_{j=1}^n d_{k, j}^{\text{loc}} + \sum_{j=1}^n d_{k, j}^{\text{off}} \leq \sum_{\ell=1}^L \mathbbm{1}_{T_\ell \in {\cal S}_{k,n}^{\text{CTS}}}(1- \alpha_\ell)D_{\ell},~n\in{\cal N},
 \end{equation}
 where $k\in{\cal K}$, and $\mathbbm{1}_{A}$ denotes the indicator function with $\mathbbm{1}_{A} =1$ if the statement $A$ is true, and $\mathbbm{1}_{A}=0$ otherwise.

 Note that the WDs need to obtain the computed results of the arrived tasks before the end of the $N$th slot. Therefore, we have the task computation deadline constraint as
 \begin{equation}\label{eq.task-deadline}
  \sum_{j=1}^N d_{k, j}^{\text{loc}} + \sum_{j=1}^N d_{k, j}^{\text{off}} = \sum_{\ell=1}^L \mathbbm{1}_{T_\ell \in {\cal S}_{k,N}^{\text{CTS}}}(1- \alpha_\ell)D_{\ell},
 \end{equation}
 where $k\in{\cal K}$. Note that $d_{k, N}^{\text{off}}=0$, since there has no time for the MEC server's remote execution at the end of the $N$th slot.

 Denote by $C_k$ the number of CPU cycles for executing one task input-bit by the local computing of WD-$k$. We consider that these CPU cycles are locally executed by WD-$k$ using an identical CPU frequency at the $n$th slot, which is determined as $f_{k, n}=\frac{C_kd_{k, n}^{\text{loc}}}{\tau}$, $\forall k\in{\cal K}$, $n\in{\cal N}$ \cite{ref1,FW18}. For WD-$k$, we assume that the CPU frequency $f_{k, n}$ is always smaller than the allowable maximum CPU frequency. Denote by $E_{k}^{\text{loc}}$ the total amount of energy consumption of WD-$k$ for local computing. Therefore, we have $E_{k}^{\text{loc}} = \sum_{n=1}^N \zeta_k C_kd_{k, n}^{\text{loc}} f_{k, n}^2= \sum_{n=1}^N\frac{\zeta_kC_k^3(d_{k, n}^{\text{loc}})^3}{\tau ^2}$, where $\zeta_k$ denotes the CPU architecture capacitance coefficient of WD-$k$.

 Let $p_{k,n}>0$, $h_{k,n}\in\mathbb{C}$, and $B_{k,n}>0$ denote the transmit power, the channel coefficient, and the system bandwidth for task offloading from WD-$k$ to the AP at the $n$th slot of Phase II, respectively. The channel state information $\{h_{k,n}\}$ is assumed to be perfectly obtained based on channel estimation methods in this letter. As WD-$k$ needs to offload a number $d_{k, n}^{\text{off}}$ of task input-bits to the MEC server, the data rate for offloading from WD-$k$ to the AP at the $n$th slot is $r_{k, n} = d_{k, n}^{\text{off}}/\tau$, where $r_{k, n} \triangleq B_{k, n}\log_2(1 + \frac{p_{k, n}|h_{k, n}|^2}{\sigma^2})$. Hence, the amount of energy consumption for WD-$k$'s task offloading in Phase~II is given by
 $E_{k}^{\text{off}} = \sum_{n=1}^{N-1}p_{k,n}\tau= \sum_{n=1}^{N-1}\frac{\tau\sigma^2(2^{{d_{k, n}^{\text{off}}}/({\tau B_{k,n}})}-1)}{|h_{k, n}|^2}$.

 As a result, the total energy consumption $E_k$ of WD-$k$ in Phase~II is expressed as $E_k=E_{k}^{\text{loc}}+E_{k}^{\text{off}}$, $\forall k\in{\cal K}$.

\subsubsection{Task Execution of MEC Server}
 The MEC server needs to execute the offloaded tasks from the $K$ WDs. Denote by $d_{n}^{\text{mec}}$ the number of task input-bits executed by the MEC server at the $n$th slot. Due to the task causality conditions, the total number of task input-bits executed by the MEC server until the $n$th slot cannot exceed those offloaded from the $K$ WDs until the previous $(n-1)$th slot. Therefore, the task causality constraints at the MEC server are expressed as
 \begin{equation}\label{eq.MEC-causality}
 \sum_{j=1}^n d_{j}^{\text{mec}} \leq \sum_{j=1}^{n-1}\sum_{k=1}^K d_{k, j}^{\text{off}},~\forall n\in{\cal N}\setminus\{N\}.
 \end{equation}
 Note that $d_1^{\text{mec}}=0$, since there exist no offloaded tasks available at the MEC server at the first slot. Again, the computation of these offloaded tasks needs to be completed before the end of the $N$th slot of Phase~II. Thus, the task computation deadline constraint at the MEC server is
 \begin{equation}\label{eq.MEC-deadline}
 \sum_{j=1}^N d_{j}^{\text{mec}} = \sum_{j=1}^{N-1}\sum_{k=1}^K d_{k, j}^{\text{off}}.
 \end{equation}
 Let $f_{n}^{\text{mec}}$ denote the CPU frequency of the MEC server at the $n$th slot, which is determined as $f_{n}^{\text{mec}}=\frac{C_0d_n^{\text{mec}}}{\tau}$. The amount of energy consumption for the MEC server to execute a total of $\sum_{n=1}^N C_0d_n^{\text{mec}}$ CPU cycles within the $N$ slots is expressed as $E^{\text{mec}} = \sum_{n=1}^N\frac{\zeta_0C_0^3(d_n^{\text{mec}})^3}{\tau^2}$.

\vspace{-0.2cm}
\subsection{Problem Formulation}
 In this letter, we are interested in minimizing the weighted-sum energy consumption of a block for the cache-enabled multiuser MEC system, subject to the MEC server's caching capacity constraint, the task causality constraints, and the task completion deadline constraints. Accordingly, by defining ${\bm x}\triangleq ( \{a_{\ell}\}_{\ell=1}^L, \{\tilde{d}^{\text{off}}_{k_o, i},\tilde{d}^{\text{mec}}_i\}_{i=1}^{N_p}, \{d_{k, n}^{\text{off}}, d_{k, n}^{\text{loc}}\}_{k\in{\cal K},n\in{\cal N}}, \{d_n^{\text{mec}}\}_{n=1}^N)$, the cache-enabled MEC design problem is formulated as
 \begin{subequations}\label{prob1}
 \begin{align}
 &(\text{P1}):~\underset{\bm x}{\text{minimize}}~ w_0 (\tilde{E}^{\text{mec}} + E^{\text{mec}}) + w_1(\tilde{E}^{\text{off}}_{k_o} + \sum_{k=1}^K E_k)\\
&\text{subject~to}~~\eqref{storage_cons1}\text{--}\eqref{eq.MEC-deadline},~\alpha_{\ell}\in\{0,1\},~\forall \ell =1,..., L \\
 &\quad\quad\quad\quad~~~ \tilde{d}^{\text{off}}_{k_o, i}\geq 0,\tilde{d}_i^{\text{mec}}\geq 0, ~\forall i=1,..., N_p\\
 &\quad\quad\quad\quad~~~d^{\text{loc}}_{k, n}\ge 0, d^{\text{off}}_{k, n}\ge 0, d_n^{\text{mec}}\geq 0,~\forall k, \forall n,
\end{align}
\end{subequations}
 where $w_0\geq 0$ and $w_1\geq 0$ denote the energy weights such that $w_0+w_1=1$. Note that (P1) is a mixed-integer nonlinear programming (MINLP) problem, which is NP-hard\cite{Boyd-Book,ref25}.

\vspace{-0.3cm}
 \section{Proposed Offline Solutions to Problem (P1)}
 In this section, we first employ BnB method to obtain the optimal offline solution to (P1), and then introduce two low-complexity schemes based on task-popularity and convex relaxation, respectively.

\subsection{Optimal Offline Solution Based on BnB Algorithm}
 The BnB method is an efficient and powerful tree-search algorithm by maintaining a provable upper and lower bound on the optimal objective value, and terminating with an $\epsilon$-optimal solution \cite{ref25}. Hence, in order to obtain the globally optimal benchmark for practical cache-enabled MEC design schemes, we employ the BnB method to solve problem (P1) in this subsection.

 To start with, we define the sets ${\cal L}_0\subseteq{\cal L}$ and ${\cal L}_1\subseteq{\cal L}$, where ${\cal L}\triangleq \{1,..., L\}$. Consider an optimization problem as
 \begin{subequations}\label{prob2}
 \begin{align*}
 \text{P}({\cal L}_0,{\cal L}_1):~ &\underset{\bm x}{\text{minimize}}~w_0 (\tilde{E}^{\text{mec}} + E^{\text{mec}}) + w_1(\tilde{E}^{\text{off}}_{k_o} + \sum_{k=1}^K E_k)\\
 &\text{subject~to}~\eqref{storage_cons1}\text{--}\eqref{eq.MEC-deadline},(\ref{prob1}\text{c}),(\ref{prob1}\text{d})\\
 &\quad\quad\quad\quad~~\alpha_{\ell}\in\{0,1\},~\forall \ell\in{\cal L}\setminus({\cal L}_0\cup{\cal L}_1),
\end{align*}
\end{subequations}
 where $\alpha_{\ell}=0$ for $\ell\in{\cal L}_0$ and $\alpha_{\ell}=1$ for $\ell\in{\cal L}_1$. If the sets satisfy ${\cal L}_0\cup{\cal L}_1\neq {\cal L}$, then P$({\cal L}_0,{\cal L}_1)$ is a mixed Boolean convex problem\cite{ref25}. Following the BnB approach, we establish a binary tree with root as P$(\emptyset,\emptyset)$, and P$({\cal L}_0,{\cal L}_1)$ corresponds to a node at depth $m$ in the tree, where a number $|{\cal L}_0|+|{\cal L}_1|=m$ of Boolean variables are specified and $0\leq m\leq L$. Specifically, we obtain a global upper bound and a global lower bound in each iteration of the BnB method, where the optimal value of problem (P1) is guaranteed to be always within the range of the global upper and lower bounds. The detailed BnB procedure is described as follows.
 \begin{itemize}
 \item{\em Bounding:} By solving P$({\cal L}_0,{\cal L}_1)$ with the Boolean variables being relaxed as continuous variables, we obtain a lower bound of the optimal value of P$({\cal L}_0,{\cal L}_1)$. By rounding $\alpha_\ell$, $\forall \ell\in{\cal L}\setminus({\cal L}_0\cup{\cal L}_0)$, to be zero or one, we obtain an upper bound of the optimal value of P$({\cal L}_0,{\cal L}_1)$.

 \item{\em Branching:} By selecting one task index $\ell\in{\cal L}\setminus({\cal L}_0\cup{\cal L}_0)$, we obtain two sub-problems as P$({\cal L}_0\cup\{\ell\},{\cal L}_1)$ and P$({\cal L}_0,{\cal L}_1\cup\{\ell\})$. Letting the Boolean variables of P$({\cal L}_0\cup\{\ell\},{\cal L}_1)$ (or P$({\cal L}_0,{\cal L}_1\cup\{\ell\})$) be relaxed and fixed, respectively, we obtain a lower and an upper bound of the optimal value of P$({\cal L}_0\cup\{\ell\},{\cal L}_1)$ (or P$({\cal L}_0,{\cal L}_1\cup\{\ell\})$). Then, we update the global lower and upper bounds.

 \item{\em Pruning:} At each iteration, we remove the nodes with lower bounds larger than the current global upper bound from the tree.
 \end{itemize}

 The proposed BnB method maintains a provable upper and lower bound on the optimal objective value, and it returns an $\epsilon$-optimal solution for problem (P1) \cite{ref25}, where $\epsilon>0$ denotes the tolerable error. Specifically, a number of ${\cal O}(2^{L+2}-1)\sqrt{N_p+KN+N}\log(\frac{(N_p+KN+N)/t^{(0)})}{\epsilon}))$ Newton iterations in the worst case is required to solve (P1), where $t^{(0)}>0$ denotes the initial barrier parameter of the interior-point method for obtaining the lower and upper bounds for each problem P$({\cal L}_0,{\cal L}_1)$\cite{Boyd-Book}, respectively. This is practically prohibited in the terms of computational complexity, especially when the task library size $L$ is large. Hence, we propose in the sequel two computationally-efficient solutions by separating caching and computation decisions.

\vspace{-0.3cm}
\subsection{Suboptimal Solution with Task-Popularity Caching Policy}
 In this subsection, we present a task-popularity caching based design scheme. First, based on the task-popularity scores of the total $L$ tasks and the MEC server's caching capacity, we determine the task cache placement decision for the task-caching phase. Next, given the cache-placement decisions, we jointly optimize the $K$ WDs' task offloading decisions and local/remote CPU frequencies within Phase II.

 For task $T_\ell$, its task-popularity score $t_\ell$ is defined as the number of occurrences in the $K$ WDs' task sequences \cite{S10,YAN2022}, i.e., $t_\ell=\sum_{k=1}^K\sum_{n=1}^N\mathbbm{1}_{s_{k,n}=T_\ell}$, where $s_{k,n}\in\mathcal{S}^{\text{CTS}}_{k,N}$. Based on the popularity scores, these $L$ tasks are ordered as $t_{\pi(1)}\geq t_{\pi(2)} \geq...\geq t_{\pi(L)}$, where $\bm\pi=[\pi(1),...,\pi(L)]^T$ is a permutation of the sequence $\{1,..., L\}$. Under the caching capacity constraint of the MEC server, we select a number of $1\leq M\leq L$ tasks with the highest-$M$ popularity scores\footnote{Note that when multiple tasks have the same popularity score, the MEC server selects the task with as large number of task input-bits as possible for energy saving, subject to the MEC server's cache capacity constraint. In the case when the equally-popular tasks have the same task input-bits, the MEC server equiprobably selects one of these tasks.}, i.e., $\{T_{\pi(1)},..., T_{\pi(M)}\}$, to be cached in the MEC server, such that $\sum_{m=1}^MD_{\pi(m)}\leq D^{\max}$ and $\sum_{m=1}^{M+1}D_{\pi(m)}> D^{\max}$. Accordingly, the sets ${\cal L}^{\text{pop}}_0=\{\pi(M+1),...,\pi(L)\}$ and ${\cal L}^{\text{pop}}_1=\{\pi(1),...,\pi(M)\}$ are determined, and we have $\alpha_i^{\text{pop}}=0$ for $i\in{\cal L}^{\text{pop}}_0$ and $\alpha_j^{\text{pop}}=1$ for $j\in{\cal L}^{\text{pop}}_1$. Next, given the determined ${\cal L}^{\text{pop}}_0$ and ${\cal L}^{\text{pop}}_1$, we solve the convex problem P(${\cal L}^{\text{pop}}_0,{\cal L}^{\text{pop}}_1$) to obtain its optimal solution $((\tilde{d}^{\text{off}}_{k_o, i})^{\text{pop}}, (\tilde{d}^{\text{mec}}_i)^{\text{pop}}, (d_{k, n}^{\text{off}})^{\text{pop}}, (d_{k, n}^{\text{loc}})^{\text{pop}}, (d_n^{\text{mec}})^{\text{pop}})$. Now, the task-popularity caching based solution for (P1) is obtained as $(\alpha_\ell^{\text{pop}}, (\tilde{d}^{\text{off}}_{k_o, i})^{\text{pop}}, (\tilde{d}^{\text{mec}}_i)^{\text{pop}}, (d_{k, n}^{\text{off}})^{\text{pop}}, (d_{k, n}^{\text{loc}})^{\text{pop}}, (d_n^{\text{mec}})^{\text{pop}})$.

\vspace{-0.3cm}
\subsection{Suboptimal Solution Based on Convex Relaxation}
 In this subsection, we present a convex relaxation based design scheme. Specifically, by relaxing the binary task cache decision variables $\{\alpha_\ell\}_{\ell=1}^L$ into continuous ones (i.e., $0\leq \alpha_\ell \leq 1$, $\forall \ell=1,...,L$), problem (P1) is transformed into a convex optimization problem, whose optimal solution can thus be efficiently obtained by off-the-shelf convex solvers, e.g., CVX toolbox\cite{Boyd-Book}. Denote by $(a_{\ell}^*, \tilde{d}^{\text{off}*}_{k_o, i}, \tilde{d}^{\text{mec}*}_i, d_{k, n}^{\text{off}*}, d_{k, n}^{\text{loc}*}, d_n^{\text{mec}*})$ the optimal solution to the convex-relaxed problem (P1). We determine the sets ${\cal L}^{\text{rel}}_0=\{\ell |0\leq \alpha_\ell^*\leq 0.5,\ell\in{\cal L}\}$ and ${\cal L}^{\text{rel}}_1=\{\ell |0.5< \alpha_\ell^*\leq 1,\ell\in{\cal L}\}$. Hence, we have $\alpha_i^\text{rel}=0$ for $i\in{\cal L}^{\text{rel}}_0$ and $\alpha_j^\text{rel}=1$ for $j\in{\cal L}^{\text{rel}}_1$, and the solution $((\tilde{d}^{\text{off}}_{k_o, i})^{\text{rel}}, (\tilde{d}^{\text{mec}}_i)^{\text{rel}}, (d_{k, n}^{\text{off}})^{\text{rel}}, (d_{k, n}^{\text{loc}})^{\text{rel}}, (d_n^{\text{mec}})^{\text{rel}})$ for Phase II.

\vspace{-0.2cm}
\section{Numerical Results}
 In this section, we evaluate the effectiveness of the proposed schemes. In simulations, we set $K=20$, $N_p=5$, $N=30$, and $\tau=0.1$~second. The CPU architecture capacitance coefficients are set as $\zeta_k=10^{-28}$ and $\zeta_0= 10^{-29}$; the number of CPU cycles for WD-$k$'s local computing and MEC server's execution of one task input-bit is $C_k=3\times10^3$ and $C_0= 10^3$ CPU-cycles/bit, $k\in\mathcal{K}$, respectively; the energy weights of the AP and the WDs are set as $w_0=0.1$ and $w_1=0.9$, respectively. Denote by $d_k\in[500,1000]$~meters~(m) the distance between WD-$k$ and the AP, where $d_k=500+\frac{500(k-1)}{K-1}$~m, $k\in{\cal K}$. We consider Rician fading channel model\cite{FW18}: $h_{k, n}=\sqrt{\frac{{\cal X_R}\Omega_0d^{-\alpha}_k}{1+{\cal X}_R}}h_0+\sqrt{\frac{\Omega_0d^{-\alpha}_k}{1+{\cal X}_R}}h$, $\forall k,n$, where ${\cal X}_R=3$ denotes Rician factor, $h_0=1$ is the line-of-sight (LoS) component, $\Omega_0=-32$~dB corresponds to the pathloss at a reference distance of one meter, $\alpha=3$ denotes the pathloss exponential, and $h \sim {\cal CN}(0,1)$ denotes the small-scale channel fading coefficient. The system channel bandwidth for WD-$k$'s task offloading is set as $B_{k, n}=2$~MHz, $k\in\mathcal{K}, n\in\mathcal{N}$. In addition, the input data $D_\ell$ of each computational task $T_\ell\in{\cal T}$ is set to follow a uniform distribution ${\cal U}(1,5)$~Kbits\cite{FW20R}, and the computational task $s_{k,n}\in{\cal T}$ of WD-$k$ at the $n$th slot of Phase II is set to follow a Zipf distribution \cite{YAN2022} with the shape parameter being 0.5.

 For comparison, we consider the following three benchmark schemes for cache-enabled multiuser MEC designs.
\begin{itemize}		
 \item{\em Joint design scheme without caching:} The AP has no task-caching functionality, which corresponds to solving (P1) by setting $a_{k, n}=0$, $\forall k, n$.

\item{\em Full offloading scheme:} Each WD executes its tasks only by offloading, which corresponds to solving (P1) by setting $d_{k, n}^{\text{loc}}=0$, $\forall k, n$.

\item{\em Full local computing scheme:} Each WD only locally executes its tasks, which corresponds to solving (P1) by setting $d_{k, n}^{\text{off}}=0$, $\forall k, n$.
\end{itemize}

 \begin{figure}
 \centering
 \includegraphics[width=2.1in]{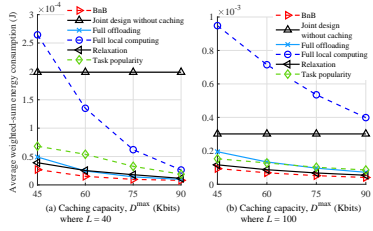}
 \caption{Performance comparison versus the MEC server's caching capacity $D^{\max}$: (a) System task set size $L=40$; (b) System task set size $L=100$.} \label{fig_VS_Dmax}
 \end{figure}

 Fig.~\ref{fig_VS_Dmax} shows the average weighted-sum energy performance versus the caching capacity $D^{\max}$, where the noise power is $\sigma^2=10^{-8}$~Watt~(W). Except for the benchmark scheme that the MEC server cannot cache tasks, the system weighted-sum energy consumption of the other five schemes decreases with $D^{\max}$. In Fig.~\ref{fig_VS_Dmax}(a), compared to the \emph{Full offloading} scheme, the proposed \emph{Relaxation} scheme achieves a closer performance to the BnB optimal scheme in the case with a small $D^{\max}$ value (e.g., $D^{\max}\leq 60$~Kbits), but it is not true with a large $D^{\max}$ value. This implies the importance of exploiting both offloading and local computing capabilities for energy saving in the case of a small caching capacity. In Fig.~\ref{fig_VS_Dmax}(a), the task-popularity based caching scheme performs inferiorly to both the \emph{Relaxation} scheme and the \emph{Full offloading} scheme, the \emph{Full local computing} scheme. In Fig.~\ref{fig_VS_Dmax}(b), the \emph{Task-popularity caching} scheme outperforms the \emph{Full offloading} scheme in the case of a small caching capacity value (e.g., $D^{\max}\leq 75$~Kbits), but it is not true in the case of a larger caching capacity value. This shows the merit of the \emph{Task-popularity caching} scheme for energy saving with a large task set size $L$. Finally, all the schemes consume more energy in Fig.~\ref{fig_VS_Dmax}(b) than that in Fig.~\ref{fig_VS_Dmax}(a). This is because the causality task set size increases with the task set size $L$.

 \begin{figure}
 \centering
 \includegraphics[width=2.1in]{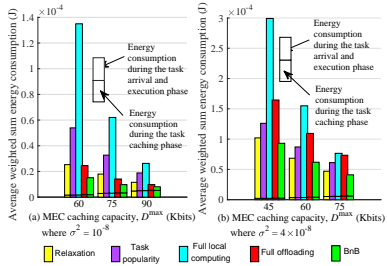}
 \caption{Performance comparison versus the MEC server's caching capacity $D^{\max}$: (a) Noise power $\sigma^2 = 10^{-8}$~W; (b) Noise power $\sigma^2 = 4\times10^{-8}$~W.} \label{fig_VS_Dmax_bar}
 \end{figure}

 Fig.~\ref{fig_VS_Dmax_bar} shows the energy consumption performance of the task caching and task arrival/execution phases, respectively, where the task set size is $L= 40$. It is observed that all the five schemes with MEC caching capability consume almost the same energy during the task caching phase, but it is not true for the task arrival/execution phase. This is because the MEC server prefers to cache computational tasks as many as possible for energy saving. In Fig.~\ref{fig_VS_Dmax_bar}, the \emph{Task-popularity caching} scheme performs inferiorly to the \emph{Relaxation} scheme, and a substantially large performance gap is observed between the \emph{BnB} and \emph{Relaxation} scheme. The \emph{Task-popularity caching} scheme outperforms the \emph{Full offloading} scheme in Fig.~\ref{fig_VS_Dmax_bar}(b), but it is not true in Fig.~\ref{fig_VS_Dmax_bar}(a). This demonstrates that the energy consumption for task offloading becomes dominant in the case with a high noise power.

\vspace{-0.3cm}
 \section{Conclusion}
 In this letter, we investigated a joint task cache placement and offloading design for cache-enabled MEC systems with dynamic task arrivals. With the objective of minimizing the system's weighted-sum energy consumption in both the task caching and task arrival/execution phases, we jointly optimized the task cache placement, the MEC server's task execution, and local computing as well as task offloading of the WDs, subject to the caching capacity, task causality, and task completion deadline constraints. We first employed the BnB method to obtain the optimal offline solution to characterize a performance lower bound for online schemes considering (partially) unknown dynamic task arrivals, and then proposed two low-complexity caching strategies based on task-popularity and convex relaxation, respectively. As a future work, it is worth investigating the robust task offloading and caching design against predicted errors of the task sequence and reinforcement learning (RL) based joint design for scenarios of partially predicable and fully unknown task-arrival sequences, respectively.


\vspace{-0.2cm}
\begin{thebibliography}{00}
\bibitem{ref1}
 Y. Mao, C. You, J. Zhang, K. Huang, and K. B. Letaief, ``A survey on mobile edge computing: The communication perspective,'' {\em IEEE Commun. Surveys Tuts.}, vol. 19, no. 4, pp. 2322--2358, 4th Quart. 2017.

\bibitem{FW18}
 F. Wang, J. Xu, and Z. Ding, ``Multi-antenna NOMA for computation offloading in multiuser mobile edge computing systems,'' {\em IEEE Trans. Commun.}, vol. 67, no. 3, pp. 2450--2463, Mar. 2019.

\bibitem{FW20}
 F. Wang, J. Xu, and S. Cui, ``Optimal energy allocation and task offloading policy for wireless powered mobile edge computing systems,'' {\em IEEE Trans. Wireless Commun.}, vol. 19, no. 4, pp. 2443--2459, Apr.~2020.

\bibitem{AL19}
H. A. Alameddine, S. Sharafeddine, S. Sebbah, S. Ayoubi, and C. Assi, ``Dynamic task offloading and scheduling for low-latency IoT services in multi-access edge computing,'' {\em IEEE J. Sel. Areas Commun.,}, vol. 37, no. 3, pp. 668--682, Mar. 2019.

\bibitem{GUO18} Y. Guo, Q. Yang, F. R. Yu, and V. C. M. Leung, ``Cache-enabled adaptive video streaming over vehicular networks: A dynamic approach,'' {\em IEEE Trans. Veh. Technol.}, vol. 67, no. 6, pp. 5445--5459, Jun. 2018.

\bibitem{YAN21} J. Yan, S. Bi, L. Duan, and Y.-J. A. Zhang, ``Pricing-driven service caching and task offloading in mobile edge computing,'' {\em IEEE Trans. Wireless Commun.}, vol. 20, No. 7, pp. 4495--4512, Jul. 2021.

\bibitem{WANG17} C. Wang, C. Liang, F. R. Yu, Q. Chen, and L. Tang, ``Computation offloading and resource allocation in wireless cellular networks with mobile edge computing,'' {\em IEEE Trans. Wireless Commun.}, vol. 16, no. 8, pp. 4924--4938, Aug. 2017.

\bibitem{XU18} J. Xu, L. Chen, and P. Zhou, ``Joint service caching and task offloading for mobile edge computing in dense networks,'' in {\em Proc. IEEE INFOCOM}, Honolulu, HI, 2018, pp. 207--215.

 \bibitem{S10}
 S. Bi, L. Huang, and Y.-J. A. Zhang, ``Joint optimization of service caching placement and computation offloading in mobile edge computing systems,'' {\em IEEE Trans. Wireless Commun.}, vol. 19, no. 7, pp. 4947--4963, Jul. 2020.

\bibitem{YAN2022}
Y. Lin, Y. Zhang, J. Li, F. Shu, and C. Li, ``Popularity-aware online task offloading for heterogeneous vehicular edge computing using contextual clustering of bandits,'' {\em IEEE Internet Things J.}, vol. 9, no. 7, pp. 5422-5433, Apr.~2022.

\bibitem{FW20R}
F. Wang, H. Xing and J. Xu, ``Real-time resource allocation for wireless powered multiuser mobile edge computing with energy and task causality,'' {\em IEEE Trans. Commun.}, vol. 68, no. 11, pp. 7140--7155, Nov.~2020.

\bibitem{Boyd-Book}
 S.~Boyd and L.~Vandenberghe, {\em Convex Optimization}. Cambridge, U.K.: Cambridge Univ. Press, 2004.

\bibitem{ref25}
 S. Boyd and J. Mattingley, ``Branch and bound methods,'' Stanford Univ., Stanford, CA, Tech. Rep., May 2011. 

\end{thebibliography}
\end{document}